\input harvmac 

\Title{\vbox{\baselineskip12pt\hbox{}}}
{\vbox{\centerline{A Sinister Extension of the Standard Model} \vskip3pt
\centerline{to $SU(3)\times SU(2)\times SU(2)\times U(1)$}}         }


\centerline{Sheldon L. Glashow}
\bigskip\centerline{Physics Department}
\centerline{Boston University}\centerline{Boston, MA 02215}

\vskip .3in

This paper describes work done in collaboration with Andy Cohen.  In our
model, ordinary fermions are accompanied by an equal number `terafermions.'
These particles are linked to ordinary quarks and leptons by an
unconventional CP'  operation, whose soft breaking in the Higgs mass sector
results in their acquiring large masses. The model leads to no detectable
strong $CP$ violating effects, produces small Dirac masses for neutrinos, and
offers a novel alternative for dark matter as electromagnetically bound
systems  made of terafermions.

\Date{05/05} 

\newsec{Introduction}

I am honored to present the closing talk at the XI Workshop on Neutrino
Telescopes, but it will not be a summary of our proceedings.  Instead, I
shall sketch some recent work~\ref\rsgac{A.G. Cohen and S.L. Glashow, {\it in
preparation.}}\ I have done with Andy Cohen.  Our model is based on the gauge
group $SU(3)\times SU(2)\times SU(2)'\times U(1)$ and involves twice as many
fermions as the standard model. The familar quarks and leptons are
accompanied by an equal number of much heavier ``terafermions.'' The model
also involves heavy versions of the weak intermediaries: $W'$ and $Z'$
bosons.  Ordinary fermions form three families, each consisting of 16
left-handed fields transforming as usual under $SU(3)\times SU(2)\times
U(1)$.  Terafermions, also forming three families of 16 left-handed fields,
transform in exactly the same way 
but under a different gauge group, $SU(3)\times
SU(2)'\times U(1)$.

We introduce an
unconventional $CP$ operation, hereafter called $CP'$, which   maps ordinary
fermions into the conventional $CP$ conjugates of their tera-equivalents, and
{\it vice versa.}  Soft  $CP'$ breaking within a simple Higgs sector
(consisting of one $SU(2)$ doublet and one $SU(2)'$ doublet) leads to large and
empirically acceptable masses for the terafermions, and for
$W'$ and $Z'$ as well.  This  `sinister' 
({\it i.e.,} `left-left symmetric')  model is somewhat akin
to (but more natural than) certain `left-right symmetric' models for which an
unconventional space-reflection operation $P'$, rather than $CP'$, links
ordinary and exotic fermions.

Regardless of the truth of our model, some its consequences may justify its
presentation here. The standard model is known to
suffer from two serious problems: the
mass hierarchy and strong $CP$. Our resolution to the latter puzzle is
subsumed under the mass hierarchy rubric, as contributions to $\overline
\theta$ from quarks are exactly cancelled by opposite contributions from much
heavier teraquarks.  A natural seesaw mechanism enables observed neutrinos to
secure tiny Dirac masses, thereby forbidding neutrinoless double beta decay.
Furthermore, we predict the existence of heavy non-standard stable quarks and
leptons.  If there is a relic abundance of these particles, they will have
formed electromagnetically bound states (`terahelium atoms') that can serve
as novel candidates for the dark matter of the universe.  \vfill\eject

 \newsec{The Model}

 Each familiar  fermion family includes  a
pair of colored quarks ({\it e.g.,} $u$ and $d$) 
with baryon number $B=1/3$, a lepton
pair ({\it e.g.,}  $e^-$ and $\nu_e$)  with lepton number $L=1$ and a neutral 
left-handed singlet ({\it e.g.,}
$n_e$) to which we assign $L=-1$. Likewise, each terafamily includes  a
pair of teraquarks ({\it e.g.,} $U$ and $D$)
with terabaryon number $B'=1/3$, a
teralepton pair ({\it e.g.,}  $E^-$ and $\nu_e'$) 
with teralepton number $L'=1$ and a
neutral singlet singlet ({\it e.g.,}  $n_e'$) with $L'=-1$. 
For reasons soon to be apparent,
 we require the  conservation  
of  $B-L$ minus its tera-equivalent,
\eqn\ef{{\cal F} \equiv (B-L) - (B'-L')\,.}
Our
 unconventional $CP'$ operation maps
each ordinary field ({\it e.g.,} $u$) 
to the conventional CP conjugate $\overline U$ of
the corresponding terafield,   and {\it vice versa} 
($U\rightarrow \overline u$).  $CP'$ {\it is assumed to be a symmetry of all
dimension-4 terms in the Lagrangian,}  but  to be softly and
severely broken by the dimension-2 mass terms of scalar mesons. Invariance
under the conventional $CP$ operation is not imposed at any level.   

The model requires two scalar Higgs multiplets.  One ($h$) transforms as an
$SU(2)$ doublet but is $SU(2)'$ invariant. It suffers Yukawa couplings to
ordinary quarks and leptons with coupling constants denoted by $\lambda$
(which compactly signifies four distinct $3\times 3$ flavor matrices related
to the masses and mixings of $Q=2/3$ quarks, $Q=-1/3$ quarks, charged leptons
and neutral leptons.)  The second Higgs multiplet $h'$ transforms as an
$SU(2)'$ doublet but is $SU(2)$ invariant.  It suffers Yukawa couplings
$\lambda'$ to teraquarks and teraleptons. Under $CP'$, $h$ is replaced by the
$CP$ conjugate of $h'$ and {\it vice versa.}

Conservation of  $\cal F$ implies that  the Yukawa couplings of
$h$ and $h'$  cannot
link fermions to terafermions. Furthermore, the $CP'$
invariance of all dimension-4 terms  relates the couplings of $h$
and $h'$:
\eqn\elambda{\lambda'=\lambda^*\,,}   
where the asterisk denotes complex conjugation.
 The vacuum expectation value of $h$  has its conventional value
$\langle h \rangle \approx 250~$GeV, but  
soft $CP'$ breaking in the Higgs mass terms results in a much larger
vev for $h'$:
\eqn\eS{\langle h'\rangle= S\langle h\rangle \gg 250\;\rm GeV\,.} 

According to Eq.\elambda, the
Yukawa couplings of fermions
and terafermions are  complex conjugates of
one another. Thus, 
each terafermion mass equals that of its lighter counterpart, but
{\it multiplied by the large factor $S$.}  The least massive
 charged terafermion is the tera-electron $E^-$, 
 a stable particle  with  mass 
$S\times$511 keV. 
 Unsuccessful searches~\ref\re{P. Achard
{\it et al.,}  Phys. Lett. B517 (2001) 75.}
 for heavy stable leptons imply  that $E^-$ 
 must be heavier than 100~GeV/$c^2$, and therefore:
\eqn\eSS{S> 2\times 10^5\,.}
Our  subsequent considerations of terafermions as dark matter
will yield a somewhat stronger constraint on  $S$.

The other charged teraleptons are unstable. They decay via $W'$ exchange at
rates that are
a factor $S$ greater than those of their ordinary counterparts.
 The same is true for the
heavier teraquarks, {\it e.g.,} $D\rightarrow U+E^-+ \overline{\nu}_e'$, 
 with the
relase of several TeV of kinetic energy. But
the tera-up quark $U$, with mass $\sim S\times$3~MeV, is stable.  With
$S=10^6$, our stable teralepton $E$ lies at $\sim 0.5$~TeV, our stable
teraquark $U$ at $\sim 3$~TeV.
These particles may be produced and detected at the LHC.

\newsec{Neutrino Masses}

Because
this Conference focusses on neutrino physics,  I shall begin by exploring the
implications of our model for that discipline.  Conservation of $\cal F$
implies that neutrinos cannot have Majorana masses, and consequently that
{\it neutrinoless double beta decay is absolutely forbidden.}  Let us see how
neutrinos naturally acquire tiny Dirac masses through a version of the seesaw
mechanism.

We have invoked
 twelve left-handed neutral
leptons. The six with ${\cal F} =-1$ are the $SU(2)$-doublet states $\nu_i$ and
the singlets $n_i'$. The six with ${\cal F}=+1$ are the $SU(2)'$ doublet states
$\nu_i'$ and the singlets $n_i$.  The Yukawa couplings of $h$ and $h'$
yield  the following contributions to neutrino masses:
\eqn\ny{\langle h\rangle\,(\nu\,\lambda_\nu\, n) + S\langle h\rangle\,
(n'\,\lambda_\nu^\dagger\,\nu')
\equiv (\nu\, m\, n) + S(n'\, m^\dagger\, \nu')}
where $\lambda_\nu$ and $m$  are  $3\times 3$ matrices and we have  used
Eq.\elambda.
(Dirac matrices are suppressed
 and $m^\dagger$ is the hermitean adjoint of $m$.)  The  eigenvalues 
of $m$ should  be
 comparable to the charged lepton masses, at least 
insofar as  up and down quark
 masses are comparable to one another.

This is not the whole story. Only one fermionic operator
 is compatible with both gauge invariance and  $\cal F$ conservation:
$(n'\,M\,n)$, where $M$ is an arbitrary  $3\times 3$ matrix 
(which is hermitean if $CP'$ is to be a symmetry of dimension-3 terms in the
 Lagrangian).
Its eigenvalues 
are expected  to be large, perhaps  comparable to a hypothetical
unification scale.  Putting this 
 mass operator together with \ny, we obtain for
the neutral lepton mass terms:
\eqn\enmm{{\cal M} = \pmatrix{\nu&n'}\,\pmatrix{0&m\cr Sm^\dagger& M\cr}\,
\pmatrix{\nu'\cr n\cr}\,,}
which  is explicitly $\cal F$ conserving and  describes two
sets of three Dirac particles. If, as we expect,
 the eigenvalues of $M$ are all much larger
than those of $m\sqrt{S}$, the seesaw
approximation is applicable.  In this  limit,
the heavy neutral leptons are made up of the states $n$ and $n'$,
with their masses   described 
by the matrix $M$.  The light neutrinos involve $\nu$ and $\nu'$, with  
their masses  described by the $3\times 3$ matrix:
\eqn\nmmmm{M_{\nu \nu'}\,\simeq\,  -S\,(m\,M^{-1}\,m^\dagger)\,.}
  With $S\sim 10^6$,  $M\sim 10^{17}$~Gev
and   $m$ plausibly chosen, we obtain  suitable neutrino masses
and mixings.  However, if $S$ is much larger than $10^6$, the eigenvalues of
$M$ would have to be  disconcertingly close to   the Planck mass. 

Corrections to the seesaw approximation are of order $|m/M|$ (or $|Sm/M|$)
for active (or sterile) neutrino states. The consequent mixings of $\nu$ with
$n'$ (and $\nu'$ with $n$) are exceedingly small and lead to no detectable
effects.  However, the tiny admixtures of active neutrino states within the
heavy neutral leptons ensure their decay in the very early universe.

In any model with Dirac neutrino masses,
 one  must examine the
contribution of  the light sterile states
($\nu_i'$ and 
$\overline{\nu}_i'$) to the expansion rate of the universe during
Big Bang nucleosynthesis. These states drop out of equilibrium 
much earlier than 
ordinary neutrinos: 
\eqn\teq{T_{eq}(\nu') = S^{4/3}\, T_{eq}(\nu)\,,} 
or $T_{eq}(\nu')\sim 300$~TeV. Many particle species  that were in thermal
equilibrium at $T_{eq}(\nu')$ 
will no longer be present
at  $T_{eq}(\nu)$.  These include the six ordinary
quarks, muons, tau leptons and their antiparticles, as well as
$W$ and $Z$ bosons and several teraspecies.
Their annihilations will reheat the conventional particle
species   relative to the 
sterile neutrinos $\nu'$. As a result,
we find~\rsgac\  the effective number of neutrino
species $N_\nu$ at  nucleosynthesis to be  about 3.15. This
result may   be
compared to  the 2-$\sigma$ limit $N_\nu \le 3.3$ deduced from astrophysical
data  with  the prior $N_\nu\ge 3$.~\ref\rbar{V. Barger {\it et al.,}
  Phys. Lett. B566 (2003) 8.}

\newsec{The Strong $CP$ Puzzle}

The quark sector of the standard model involves two independent $CP$
violating parameters: the Kobayashi-Maskawa phase $\delta$ and the strong
$CP$ parameter $\overline\theta$.  The puzzle is to explain why $\delta$ is
order unity whereas $\overline\theta<10^{-10}$.  In our model, the
$CP$-violating operator $G\tilde G$ is forbidden by our $CP'$ symmetry.
The complex Yukawa couplings of $h$ and $h'$ do not contribute to
$\overline\theta$ because the teraquark  mass matrices are proportional
to the complex conjugates of the quark mass matrices. Thus, their
contributions to $\overline\theta$ cancel one another. It follows that
$\overline\theta$ vanishes in tree approximation.

A finite and small  value for  $\overline\theta$ is 
generated by radiative corrections. These  consist only of
multi-loop diagrams such as  are present in the standard model: finite
six-loop terms and divergent seven-loop terms. In our case, the divergence
is replaced by the  factor $\ln S$. It follows that strong
$CP$-violating   effects
are entirely neglible in our model, for which   electric
dipole moments of elementary particles are far too small  to be detected.

Our solution  to  the strong $CP$ puzzle is a particular
realization  of 
the Nelson-Barr mechanism~\ref\rnb{A. Nelson, Phys. Lett B136 (1984)387\semi
S. Barr, Phys. Rev. Lett 53 (1984) 329.}, with the UV sector (teraquarks)
linked to the 
low-energy sector by a softly broken discrete symmetry. In effect,
we have replaced  the strong $CP$ puzzle by a 
novel fermion-terafermion  mass hierarchy.

\newsec{Relic Terahelium as Dark Matter?}

There are two
stable terafermions in  our model aside from the right-handed neutrino
 states 
 $\overline \nu'$. These are
 the tera-up quark $U$ and the tera-electron $E$. 
Were  there  a terabaryon asymmetry in the universe akin to the baryon
asymmetry, 
 the alleged dark matter of the universe could consist of relic
terafermions. In our discussion of this possibility, we make two technical
assumptions: that the net electric charge of relic fermions and 
 of relic
terafermions each vanish, and that both the baryon and terabaryon 
excesses are 
positive. One of the less attractive features of our model is that
each of these relic abundances must be adjusted to yield  the
known  values of the mean densities of baryons
$\Omega_b$   and dark matter $\Omega_d$ in the 
universe~\ref\rwmap{{\it E.g.,} C.L. Bennett {\it et al.,} ApJS 148 (2003)
 1.}:
\eqn\eomega{\Omega_b\simeq 0.044\quad\ {\rm and }\quad
\Omega_d\simeq 0.224\,.}

What happens to relic 
$U$'s and $E$'s in the early universe?  
The QCD force among  massive $U$'s is both strong and Coulombic. Thus,
the
binding energy of a color triplet $UU$ diquark is $\sim \alpha_s M$, or
hundreds of GeV. That of the color singlet $UUU$  is several times
larger. Thus, exothermic processes such as:
\eqn\euug{U+U\rightarrow (UU) + g\quad\ {\rm and}\quad\ 
U+ (UU)\rightarrow (UUU) + g}
where $g$ is a gluon, can proceed 
irreversibly at temperatures far above the quark-hadron
transition.  However, the cross-sections for these reactions are small, so
that not all the $U$'s are aggregated by this means.

Once the temperature becomes low enough
for conventional hadrons  to form, the unaggregated 
 $U$'s will bind  with the more abundant $u$ and
 $d$ quarks to form super-heavy hadrons
 such as  $Uud$ and $U\overline d$.
At that point,  exothermic  exchange  reactions such as:
\eqn\euu{\eqalign{Uud+Uud\rightarrow UUd +uud\,,  \quad \phantom{\rightarrow}& 
\quad U\overline d +Uud\rightarrow UUd+ u\overline d\quad {\rm and}\cr
\quad U\overline d+U\overline d\;\rightarrow&\; UUd+ \overline{ddd}\cr}
}
rapidly ensue. Cross-sections for 
 exchange reactions such as \euu\  are  much larger than
 those for reactions \euug. They are
 comparable to the 
cross-sectional areas  of the reactants, which are of order tens of millibarns.
The aggregation of $U$ quarks will continue via  further 
highly exothermic exchange reactions
such as:
\eqn\euuu{\eqalign{
UUd+ Uud\rightarrow UUU + udd\,, \quad \phantom{\rightarrow}& 
\quad 
UUd+UUd\rightarrow UUU + Udd \quad {\rm and}\cr
 UUd+U\overline d\;\rightarrow&\; UUU+d\overline{d}\,.\cr}}

Our estimates~\rsgac\  show  that reactions \euu\ and \euuu\ will 
efficiently convert
almost all relic teraquarks
 into bound $UUU$ states prior to nucleosynthesis.
 These tiny and tightly bound doubly-charged particles,
the tera-equivalents of
$\Delta^{++}$,  have spin 3/2   and mass $\sim 10$~TeV. 
Using  Eq.\eomega\ and $\eta_b\simeq 6\times 10^{-10}$ for the baryon to
photon ratio,  
we find:
\eqn\etab{\eta_{B'}\approx 3\times 10^{-13}\, S_6}
(with $S_6=10^6\,S$)
for the  tera-$\Delta^{++}$  to photon ratio 
once  the exchange reactions  have done their job.

What about the tera-electrons, of which there  are twice as many 
as  tera-$\Delta$'s?  These particles   efficiently
recombine
 with  protons once the temperature falls  below  
the $Ep$ binding energy of $\sim 25$~keV:
\eqn\erc{E^- + p\rightarrow (E^-p) +\gamma\,.}
These $E$-onic atoms
thereupon  engage in the following
 exothermic exchange reactions:
\eqn\euuue{UUU + Ep\rightarrow (UUUE) + p\quad{\rm and}\quad
(UUUE) + Ep \rightarrow (UUUEE)+p\,.}
The relevant cross-sections for \euuue\ are roughly equal to
 the cross-sectional areas of the
 $Ep$ atoms, whose radii are $\sim 3\times 10^{-12}$~cm.
Thus they are order of barns!
 These reactions   should proceed to completion because the 
energy release is far greater than the temperature.
As a result,
virtually all relic terafermions are expected to form
tiny and electrically neutral
`terahelium atoms,' with $(UUU)^{++}$ as nuclei and two bound
tera-electrons forming a closed shell.
  These are our candidates for the dark matter of the
universe.

To determine whether tera-helium atoms are 
plausible candidates for dark matter, 
 we have estimated  their interaction cross sections with atomic
nuclei~\rsgac.  There are three distinct  mechanisms for terahelium-nuclear 
scattering:

\smallskip
\noindent
$\bullet$ via  the    chromoelectric polarizability of the teranucleus
    $UUU$,\hfill\break
$\bullet$  via the electric polarizability of the tera-atom, or\hfill\break
$\bullet$ via the magnetic moment of the teraatom, which is exclusively that
    of its $UUU$ nucleus because the spins of its two  $E$'s cancel.
\smallskip

We find  the latter mechanism to dominate: the interaction between the
nuclear charge and the magnetic moment of tera-helium.
Our rough  estimate
of the tera-helium--nuclear cross-section  implies
that  $S > 10^6$ (or equivalently,  that the terahelium mass must exceed
$10$~TeV) if  our model is
to be consistent with  the negative results of current
dark matter 
searches~\ref\rdms{{\it E.g.,} A. Benoit {\it et al.,} Phys. Lett. B545
(2002) 43\semi D.S. Akerib {\it et al.,} Phys. Rev. D68 (2003) 082002.}. 
 An improvement of the sensitivity
of these experiments by one or two  orders of magnitude
could  exclude (or support) the notion of tera-helium as dark matter.

Our proposal for the nature of dark matter 
  faces another  potentially serious
obstacle.  We showed  that almost all relic terafermions end up as
electrically neutral terahelium, but almost all may not be good enough.
Remnant tera-electrons, neutral (Ep) atoms
 or exotic  hadrons such as $Uud$, if present on earth
or in cosmic rays, may 
combine with earthly elements to form super-heavy
isotopes, the abundance of which is very highly
 constrained~\ref\rshi{{\it E.g.,} P.F. Smith {\it et al.,} Nucl. Phys. B206
 (1982) 333\semi  T.K. Hemmick {\it et al.,} Phys. Rev. D41 (1990) 2074\semi
P. Verkerk {\it et al.,} Phys. Rev. Lett. 68 (1992) 1116.}.
If our model is
to survive, these remnants must have been dealt with by one or more of the
  following mechanisms:
 by being more completely aggregated into
tera-helium during structure formation, or by being gravitationally
concentrated within stars because  of their great masses, or by being
processed during nucleosynthesis into superheavy
elements other those  
for which sensitive searches have been carried out.

\bigskip {\bf Acknowledgement:} SLG was supported in part by the National
Science Foundation under grant number NSF-PHY-0099529 and AGC by the
Department of Energy under grant number DE-FG02-91ER-40676.

\listrefs

\parindent=20pt

\bye